\documentclass{article}

\usepackage{epsfig}

\begin{document}

\title{Mapping dynamical systems onto complex networks}

\author{Ernesto P. Borges$^a$, 
 Daniel O. Cajueiro$^b$ 
 and Roberto F. S. Andrade$^c$\thanks{\tt randrade@ufba.br}
\\ $ $ \\
{\normalsize $^a$Escola Polit\'ecnica, Universidade Federal da Bahia,} \\
{\normalsize R. Aristides Novis, 2, 40210-630 Salvador--BA, Brazil}
\\
{\normalsize $^b$Departamento de Economia, Universidade Cat\'{o}lica de Bras\'{i}lia} \\
{\normalsize 70790-160 Bras\'{i}lia, Brazil}
\\
{\normalsize $^c$Instituto de F\'{i}sica, Universidade Federal da Bahia,} \\
{\normalsize 40210-340 Salvador, Brazil} 
}

\date{}

\maketitle

\begin{abstract}
A procedure to characterize chaotic dynamical systems with concepts of
complex networks is pursued, in which a dynamical system is mapped onto a
network. The nodes represent the regions of space visited by the system,
while edges represent the transitions between these regions. Parameters
used to quantify the properties of complex networks, including those related 
to higher order neighborhoods, are used in the analysis. The methodology is 
tested for the logistic map, focusing the onset of chaos and chaotic regimes.
It is found that the corresponding networks show distinct features, which are 
associated to the particular type of dynamics that have generated them.
\end{abstract}
\begin{quote}
\small 
89.75.Fb: Structures and organization in complex systems \\
89.75.Hc: Networks and genealogical trees   \\
02.10.Ox: Combinatorics; graph theory
\end{quote}

\section{Introduction}
\label{sec:intro}

Chaotic dynamical systems are characterized by several measures that
quantify how irregular (despite deterministic) the trajectories are. The
set of Lyapunov exponents~\cite{eckrue85} provides a measure of the
dependence on the initial conditions of their trajectories, while
information theory may be used to characterize a dynamical system in terms
of the production of entropy. Actually, a dynamical system with a chaotic
behavior is regarded as a realization of Shannon's concept of an ergodic
information source~\cite{gal68}, as the Kolmogorov-Sinai entropy is
equated~\cite{pes77} to the sum of the positive Lyapunov exponents.
Further measures to describe a chaotic system include fractal dimensions
and singularity spectra~\cite{henpro83,mensre89}. This formalism applies
only where the system has, at least, one positive Lyapunov exponent.
However, several authors have studied situations in which
the largest Lyapunov exponent 
vanishes, but the distance between
neighbouring trajectories increases in time according to a power law
\cite{tsa97,rob04,rob05,rob06}. These situations are usually found at the
onset of chaos, where an infinitesimal change of a control parameter 
drives the system into either a regular or a chaotic regime. These
investigations uncovered some of its features
related to the sensitivity to the initial conditions ~\cite{tsa97},
entropy production per unit time ~\cite{lat00}, multifractal geometry of
the attractor ~\cite{lyrtsa98}, relaxation to the system attractor
~\cite{bor02} 
and multifractal dynamics at the onset of chaos ~\cite{mayoral05}.

Recently, the investigation of complex networks has set up a new framework
for the analysis of systems with a large number of degrees of freedom.
Within it, one has access to the properties of the topological structure
underlying the mutual interactions among the system constituents. This
approach has been applied to a large variety of actual systems, ranging
from social interactions, biological nature, information and electrical
distribution \cite{albbar02,lat06}.

In this work, we define a mapping of dynamical systems onto a network.
This way, the network properties can be used to display new features for
the characterization of the system's trajectory. The network nodes 
correspond to coarse-grained regions (cells) of the phase space visited by
the trajectory. Two nodes $r$ and $s$ are linked if, during the time
evolution, the trajectory jumps from cell $r$ to cell $s$. Although this
naturally offers a construction procedure for a directed network, we
consider here undirected networks. In this approach, we are able to find
novel geometric and topological properties of the phase space through the
measures that have been recently developed to characterize complex
networks. The dynamical system, defined on the time domain, is mapped into
a node domain, which represents regions of the phase space visited by the
trajectory. As it is based on the division of phase space into boxes, this
process follows similar construction procedures considered in the
evaluation of the fractal dimension of attractors and the Kolmogorov-Sinai
entropy. 

It worths mentioning that some previous works have put together
ideas of dynamical systems and complex networks. As we shall see, 
our approach is rather distinct, as the previous contributions consider
synchronization under the assumption of certain regularity in the
connection topology~\cite{pik01,atajos04}.
The use of a network to represent the phase space evolution of discrete time 
dynamical systems has been suggested in \cite{stephan}. 

We restrict ourselves to the analysis of the quadratic map
\cite{collet80} described by
\begin{equation}\label{eq:logistic}
x_{t+1}=1-a x_{t}^{2}\;\;(t=0,1,2,3\cdots)
\end{equation}
where $x_{t}\in [-1,1]$ and $a\in [0,2]$. Eq.~(\ref{eq:logistic}) leads to a
variety of distinct dynamical situations, the properties of which are
expected to be manifested in the networks they originate. In particular,
we explore those regimes at the onset of chaos that proceed through a 
bifurcation cascade as well as those close to an intermittency transition
and also the full developed chaos.

\section{Network characterization}
\label{sec:network}

An indirected network $R$ is defined only by the number $N$ of nodes and
links $L$, represented by an assembly of unordered pairs $S_R{(r,s)}$,
$r,s\leq N$, indicating which pairs of nodes are directly connected. This
information provides a full description of the network, leading to the
computation of the average number of links per node $\langle k\rangle$,
the average clustering coefficient $C$, the mean minimal distance among
the nodes $\langle d\rangle$, the diameter $D$, the probability
distribution of nodes with $k$ links $p(k)$. Other measures, like the
assortativity degree \cite{new02} and the distribution of individual node
clustering coefficients $C(k)$ with respect to its degree $k$ have also
been introduced, but they are not explored here. 

$R$ can be described by its adjacency matrix $M(R)$. This is not the most
concise representation of a network, but it opens the possibility to  the
evaluation of its spectral properties and, as recently indicated, the
higher order neighborhoods $R_{\ell}$, $\ell=1,2,..., D$ \cite{and06}.
This is done in a straightforward way, by means of the set of matrices
$\{M_{\ell}\}$, so defined that $(M_{\ell})_{r,s}=1$ only if the shortest
distance along the network between the nodes $r$ and $s$ is $\ell$.
Otherwise, $(M_{\ell})_{r,s}=0$. Although the whole information on the
network is entailed in $S_R{(r,s)}$ or in $M(R)$, each $M_{\ell}$
condenses information on $R$ that is extracted from $M(R)$ within the
quoted framework. This formalism is also consistent with the recently
proposed procedure to evaluate the fractal dimension of the network
$d_{F,R}$ \cite{son05}, as it naturally leads to the set $\{N_{\ell}\}$
required for this evaluation. Here, each $N_{\ell}$ counts the number of
pairs of nodes which are $\ell$ steps apart.

Within this framework, we consider that each node is the only zeroth order
neighbor of itself, and define $M_0=I$, where $I$ indicates the identity
matrix. Also, we assume that $M_1=M$. Since the matrix elements of $M(R)$
take only the values 0 and 1, the other matrices $M_{\ell}$ of the set are
recursively evaluated with the help of Boolean operations \cite{and06}. We
shall make further use of a matrix that condenses all information in
$\{M_{\ell}(R)\}$. As just discussed, given any two nodes $r$ and $s$, it
is clear that $(M_\ell)_{rs}=1$ for just one value of $\ell$. So, if we
define a matrix
\begin{equation}\label{eq:Matrix}
\widehat{M}=\sum_{j=0}^{\ell_{max}}j M_j,
\end{equation}
it directly informs how many steps apart any two nodes in the network are.
Also, it is possible to use the information in $\widehat{M}$ to
graphically display the structure of a network with the help of color or
grey scale plots.

It should also be mentioned that this framework opens the door for a finer
characterization of the network, if we consider each $R_\ell$ as an
independent network. Thus, several of the properties quoted above used to
characterize $R$ can be also evaluated for the evaluation of $R_{\ell}$.
This is explored in the next section specifically for the the degree
distribution and clustering coefficient $\langle k\rangle_\ell$ and
$C(\ell)$.

\section{An algorithm to map a dynamical system into a network}
\label{sec:algorithm}

To map a dynamical system into a network $R$, we use the framework of the
algorithm introduced in~\cite{blo90}, to efficiently evaluate the
generalized fractal dimensions of fractal structures by the box counting
method.

Let us consider a dynamical system with $m$ variables. Without loss of
generality, one can consider a set of points $\mathcal{Z}\subset \Re^m$
consisting of the vectors $z(i)$, $i=1,\cdots,T$, $T \gg 1$, which
represent the coordinates of the dynamical system. The components of these
vectors, $z_\delta(i)$, $\delta=1,\cdots,m$, are assumed to belong to the
interval $[0,1)$. We define a graining in phase space by dividing each
phase space axis into $W$ equally sized disjoint intervals, so that the
whole phase is spanned by a set of $W^m$ boxes. This represents also the
maximal possible number of nodes in a network, as {\em e.g.}, in the case
of an ergodic system. Of course, the choice of $W$ defines the graining,
and the size of the region represented by a node. In the next section we
investigate the effect of $W$ on the obtained networks.

Based on ~\cite{blo90}, each point $z(i)$ of the trajectory is mapped into
a node of $R$ according to
\begin{equation}\label{eq2}
n(i)=\sum_{\delta=1}^{m} W^{\delta-1}\mathrm{floor}(W
z_\delta(i))\mathrm{,}
\end{equation}
where $\mathrm{floor}(x)$ is a function that evaluates the largest integer
less than $x$. 
Actually, this is just a simple way to divide the region
$[0,1)^m$ in equal parts. Thus, the nodes of the so constructed network
represent a box in the coarse grained phase space of the system. After the
mapping is complete, the boxes that were not visited by the trajectory are
eliminated from the network, as they constitute nodes with zero degree
($k=0$), which do not provide any useful information on the dynamical
system. The edges are built as described in the following procedure. Let
$z(i)$ and $z(i+1)$ be two consecutive points of the dynamical system.
Suppose that these points were previously mapped into the nodes $n(r)$ and
$n(s)$, where $0<r,s\le W^m$. Thus, one assumes that there is an edge
connecting $n(r)$ and $n(s)$. Here, it is considered that there is only
one edge linking $n(r)$ to $n(s)$ if $r\ne s$, {\em i.e.} self-links are
not computed.

This procedure can lead to directed and weighted networks; however, in
this work, we focus on the most simple situation of undirected and
un-weighted networks, once our main purpose here is to address the
problem, and show that we can extract useful information from it.

\section{Results}
\label{sec:results}

We concentrate our investigation on three distinct regions of $a$,
$a=a_c=1.40115518909...$, $a\in[1.749,1.75)$ and $a=2$, which correspond,
respectively, to the first period doubling transition, the region close to
the tangent bifurcation to the period-three window, and the full developed
chaotic state. Representative networks to the distinct chaotic attractors
are generated for distinct values of the graining $W$. We only consider
trajectories that start on the attractor, in order to avoid spurious nodes
(visited only once) that depend on initial conditions. For a fixed value
$W$, the network grows as the trajectory evolves in the the phase space
with respect to the number of iteration steps $t$. There is no a priori
criterion to decide the time $t_F$ after which the network is complete. In
this work we have followed how $N$ and $L$ increase with $t$, for a given
$W$. We define $t_F$ as the smallest value of $t$ for which
$N(t_F)=N(2t_F)$ and $L(t_F)=L(2t_F)$.

\begin{figure}[tbp]
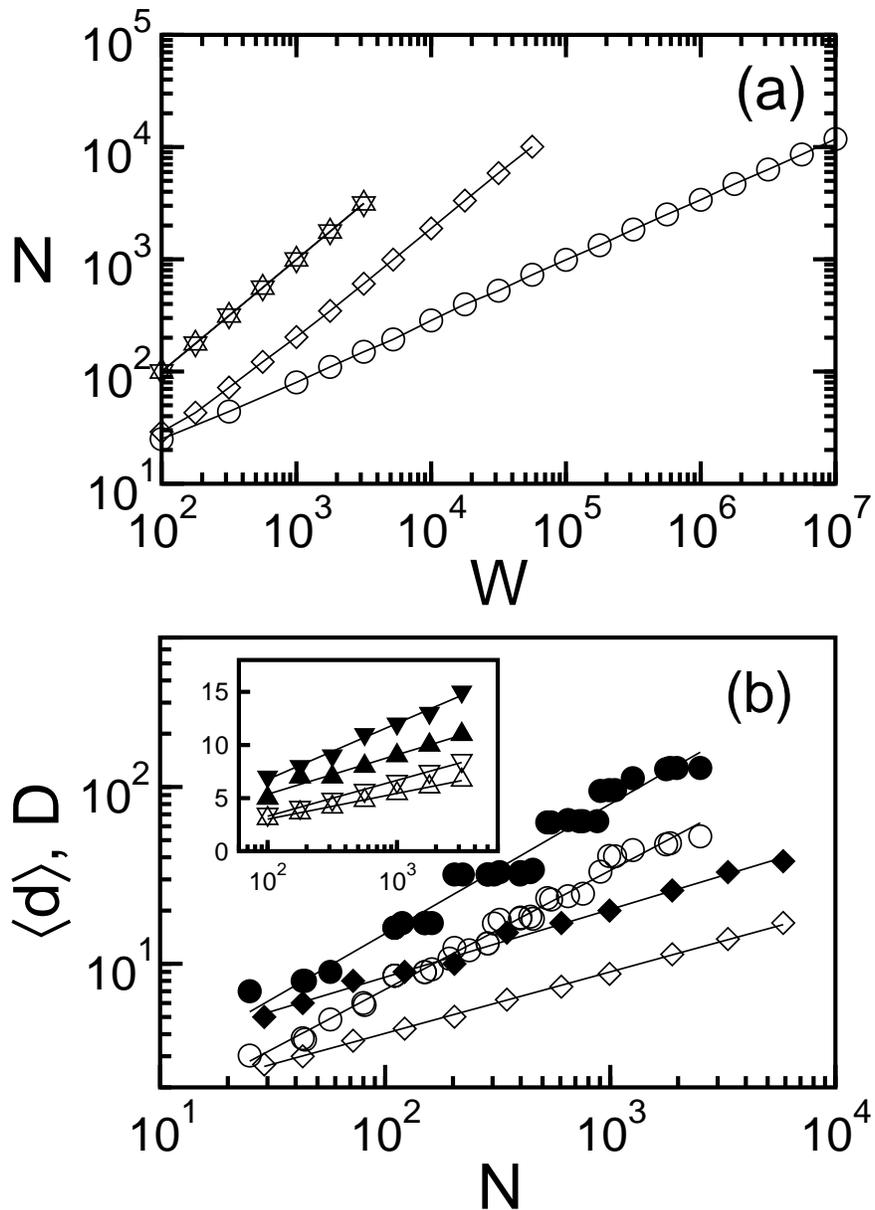

\begin{center}
\epsfig{figure=n-w.eps,width=0.95\columnwidth,keepaspectratio,clip=}
\epsfig{figure=d-w.eps,width=0.95\columnwidth,keepaspectratio,clip=}
\end{center}
\caption{(a)~Dependence of $N$ with respect to $W$ for
$a=a_c$ (circles), $a=a_c+10^{-3}$ (diamonds), $a=1.749999$ (down
triangles), and $a=2$ (up triangles). 
The the same convention is used in all other figures.
The distinct slopes indicate the values of $d_{F,A}$.
(b)~Power law dependence of $\langle d\rangle$ (hollow symbols) and $D$
(solid symbols) with respect to $N$ for $a=a_c$ and $a=a_c+10^{-3}$. The
inset shows logarithmic dependence among the same quantities when
$a=1.749999$ and $a=2$.}
\label{fig:netparameters}
\end{figure}

First let us discuss the effect of $W$ on $N$ and $L$. For the purpose of
presenting a full neighborhood analysis of the networks, we have selected
here values of $W$ that lead to the maximal number $\approx$ 10000 nodes 
in the network. Such choice for $W$ clearly depends on $a$.
Indeed, due to the strategy adopted for the construction of the networks,
$N$ grows with $W$ according to a power law mediated by the fractal
dimension of the attractor $d_{F,A}$. This is shown in
Fig.~\ref{fig:netparameters}a, where we draw points for $a=a_c$,
$a_c+10^{-3}$, $1.749999$ and 2. For $a_c$ we measure the slope
$0.54\dots$, which agrees with the known value of $d_{F,A}$ of the period
doubling  attractor. For all other cases, the slope is 1 within 2\%
accuracy, even for $a=a_c+10^{-3}$, which lies already in the chaotic
regime. This is in accordance with the fact that $d_{F,A}$ changes in a
discontinuous way at $a=a_c$. When $a$ corresponds to a periodic solution,
the network becomes finite, so that $N$ and $L$ do not depend on $W$, 
provided this parameter is large enough.

Although we are primarily interested in the properties of the complete
network, we can also follow the dependency of $N$ and $L$ with respect to
$t$, for a fixed value of $W$. Assuming $N\sim L^z$, this defines a
dynamical exponent $z$ in the early stages of the network evolution. The
analyzed data indicate that $z\simeq 1$ for all values of $a$.
Nevertheless, we find that, in the immediate chaotic neighborhood of
$a=1.75$,
the laminar phases in the intermittent regime traps the trajectory
for long intervals, demanding large time of integration
to complete the network.

Now we discuss the network properties, following the methodology and
parameters indicated before. When appropriate, we extend our discussion to
properties of higher order neighborhoods in the network. We find distinct
network structures when we consider the chaotic regime or the onset of
chaos. Regarding the mean minimal distance $\langle d\rangle$ and diameter
$D$, we find that, for the chaotic regime, they grow with respect to $W$
(and $N$), in a logarithmic way, similarly to small-world networks
\cite{watts98}. If we write $\langle d\rangle=\alpha \log_{10} W$, we find
$\alpha\simeq2.4$ and 3.7, respectively, for $a=2$ and
$a\in[1.749,1.749999]$, as illustrated in the inset of
Fig.~\ref{fig:netparameters}b. As $D$ only assumes integer values, it is
possible to expect a similar behavior only in an approximate way, {\em
e.g.}, equally sized steps in $D\times \log_{10} W$ plots. Thus, assuming
$D=\beta \log_{10} W$, we find $\beta\simeq4$ for $a=2$ with very good
accuracy. In the interval $[1.749,1.749999]$, we noticed the presence of
fluctuations in the size of the steps, which increases when we approach
the threshold $a=1.75$. The results for $a=a_c$ have a completely distinct
behavior: $\langle d\rangle$ and $D$ increase as power
laws with respect to $W$, as illustrated in the main panel of
Fig.~\ref{fig:netparameters}b. For $a=a_c+10^{-3}$, the same dependence
prevails. The exponents obtained for $\langle d\rangle$ and $D$ are,
respectively, $0.67$ and $0.73$ for $a_c$, and $0.35$ and $0.38$ for
$a=a_c+10^{-3}$. $\langle d \rangle$ and $D$ increase smoothly for
$a=a_c+10^{-3}$ but at $a=a_c$, the growth of the two quantities present
discontinuities and steps. Results in Fig.~\ref{fig:netparameters}b indicate
that, unlike the properties of the attractor, reflected by the rash change
in the value of $d_{F,A}$, the network properties change slowly when the
chaotic regime is reached.

\begin{figure}[tbp]
\begin{center}
\epsfig{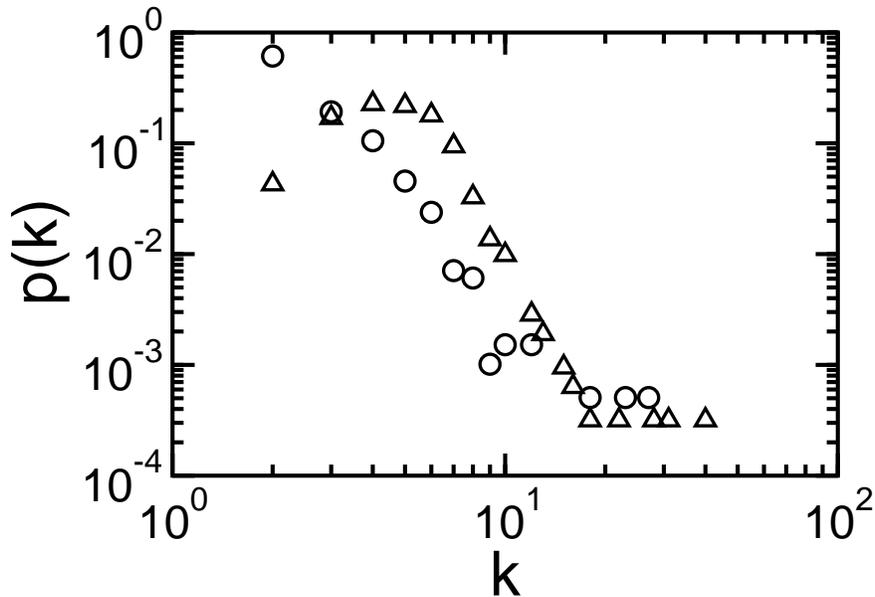}
\end{center}
\caption{
Degree distribution of nodes at $a_c$ (circles) 
and in the $a=2$ chaotic regime, (up triangles).
Behavior close to tangent bifurcation, $a=1.75 - 10^{-4}$, (not shown) 
is quite similar to $a=2$.} 
\label{fig:degree}
\end{figure}

We have evaluated the distribution $p(k)$ vs. $k$ in all different
regimes. 
We can see in Fig.~\ref{fig:degree} that, for $a=a_c$, $k$ does not reach
large values ($k_{max}\simeq 30$), so that it is not possible to identify 
a power law decay in this range.
For $a=2$ (and also $a=1.749999$) nodes with larger values of $k$ can be found,
but $p(k)$ does not follow either a power law.

\begin{figure}[tbp]
\begin{center}
\epsfig{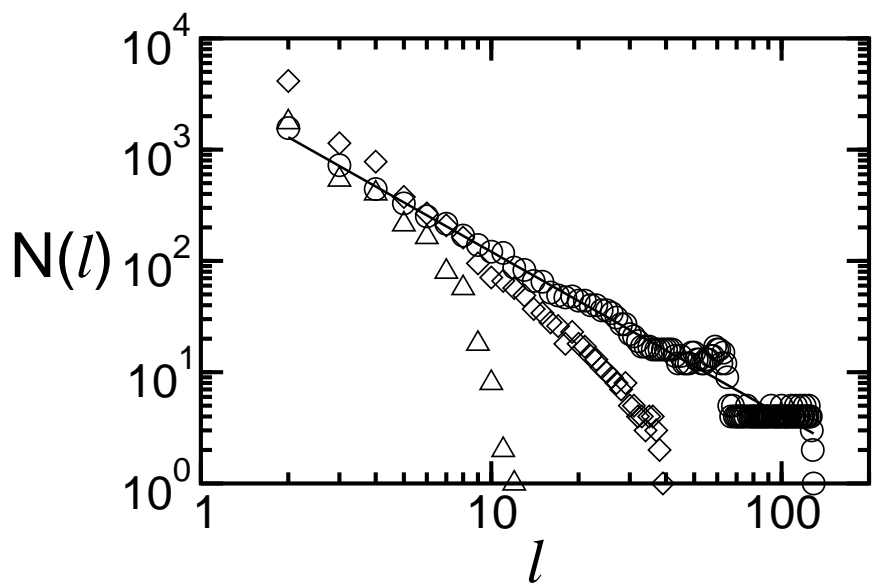}
\end{center}
\caption{Clear power law behavior for $N(\ell) \times \ell$ when $a=a_c$,
with $d_{F,R}=1.47$. Finite size effects blur this dependence when
$\ell\simeq D$. In the chaotic regime, for $a=2$, $d_{F,R}$ can
hardly be evaluated. For $a=a_c+10^{-3}$, the points illustrate a slow
transition between the two regimes.}
\label{fig:fractaldim}
\end{figure}

Such distinct behavior is also explicit when we analyze the fractal
dimension of the network $d_{F,R}$ \cite{son05}. Fig.~\ref{fig:fractaldim}
shows that the
$a=a_c$ networks have a well defined scaling behavior, which extends over
more than two decades, in a quite precise way. On the other hand, a
fractal dimension for the networks in the chaotic regime is not
evident. First, the small values of $D$ reduces the region of possible
scaling behavior. Then, we clearly observe deviations to the expected
power-law regime.

Regarding the clustering coefficient, we obtain $C\equiv0$ for $a=a_c$,
indicating the complete absence of triangles in the network. For
$a=1.749999$ and $a=2$, we find that $C$ decays with $N$ according to a
power law with exponent $\approx$ 0.95, what is not so close to the value
$0.75$ observed for the Albert-Barabasi scale-free network
\cite{albbar02}. However, its small values indicate that the network has
only a small number of triangles.

\begin{figure}[tbp]
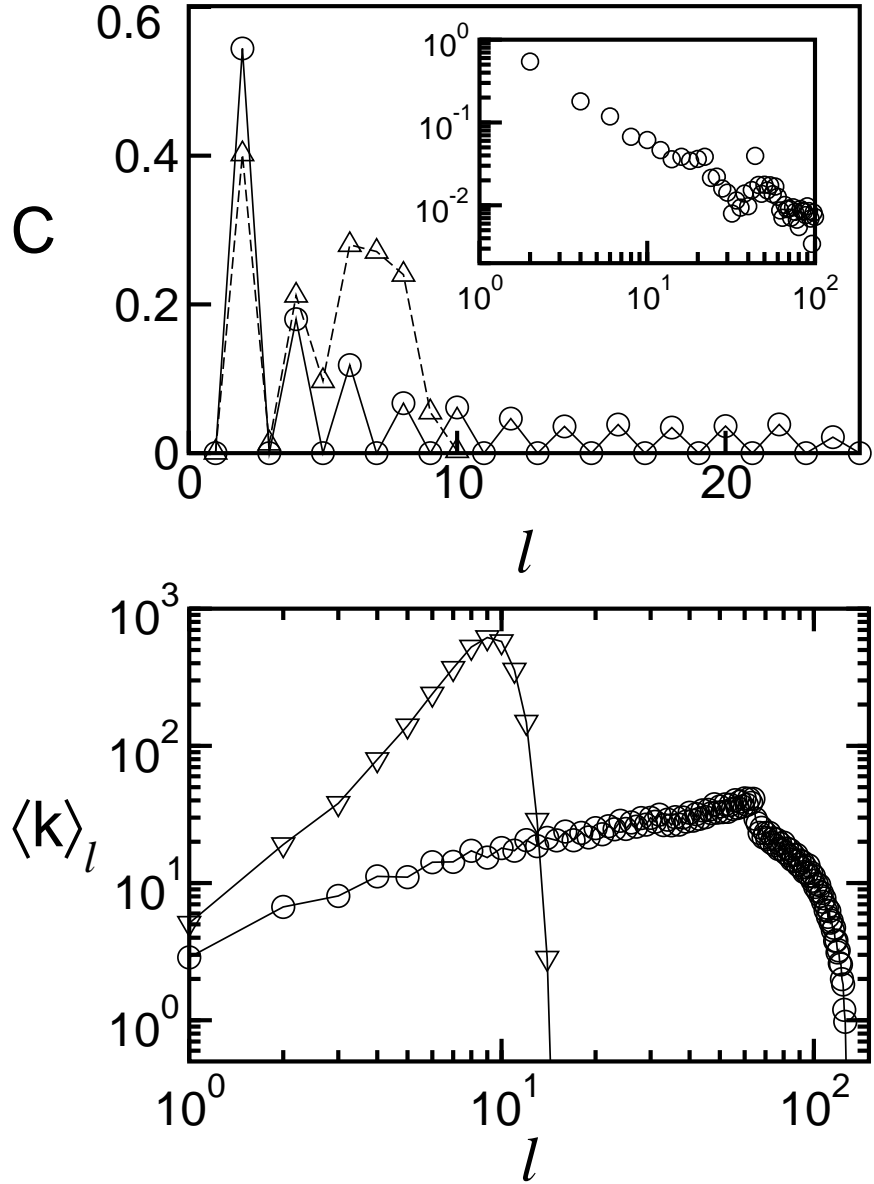

\begin{center}
\epsfig{figure=clustering.eps,width=0.95\columnwidth,keepaspectratio,clip=}
\epsfig{figure=averagedegree.eps,width=0.95\columnwidth,keepaspectratio,clip=}
\end{center}
\caption{(a)~Behavior of $C(\ell)$ with respect to $\ell$ for chaotic regime
and at the onset of chaos.
The inset shows a peculiar power law behavior $C(\ell)$ 
at the onset of chaos.
(b)~Behavior of $\langle k\rangle_{\ell} \times
\ell$. 
The onset of chaos exhibits a slow increasing
value of $\langle k\rangle_{\ell}$ over a large interval of $\ell$,
interrupted by finite size effects already present in 
Fig.~ \protect\ref{fig:fractaldim}.
In opposition to this picture, chaotic regime shows a sharp increase
of $\langle k\rangle_{\ell}$.}
\label{fig:neighborhood}
\end{figure}

Other features of the network may be drawn if we consider the clustering
coefficient of higher order neighborhoods \cite{and06}. To obtain a
clearer picture of this analysis consider, for instance, the regular
Cayley tree with $\ell=2$ network for the complete regular three neighbors
Cayley tree. The $\ell=2$ network  is formed by triangles, much like the
Husimi cactuses and, correspondingly, a large value for $C(\ell=2)$. The
following odd and even numbered neighborhoods are characterized,
respectively, by values of $C(\ell)=0$ and $C(\ell)>0$, whereby the values
of $C(\ell)$ for a subset of even neighborhoods decrease in a monotonic
way. A similar situation is found for the networks investigated here. In
Fig.~\ref{fig:neighborhood}a we summarize the sequence of $C(\ell)$ for the
three situations under investigation. We see that, for $a=2$ and
$a=1.4999$, the oscillatory behavior lasts only until $\ell=5$ and 10,
respectively. Also we note that the odd numbered $C(\ell)$ increases until
it reaches values as high as those of the even numbered neighborhoods. On
the other hand, the $a=a_c$ network has $C(\ell)=0$ for all odd numbered
neighborhood. The inset shows that the $C(\ell$=even number$)$ decays with
$\ell$ according to a power law, with exponent $\alpha\simeq 1$.

We have also analyzed the average degree $\langle k\rangle_{\ell}$ as a
function of the neighborhood $\ell$. Here again we find that the behavior
of the chaotic regime and the onset of chaos have distinct features, as
exhibited in the Fig.~\ref{fig:neighborhood}b.

\begin{figure}[tbp]
\begin{center}
\epsfig{figure=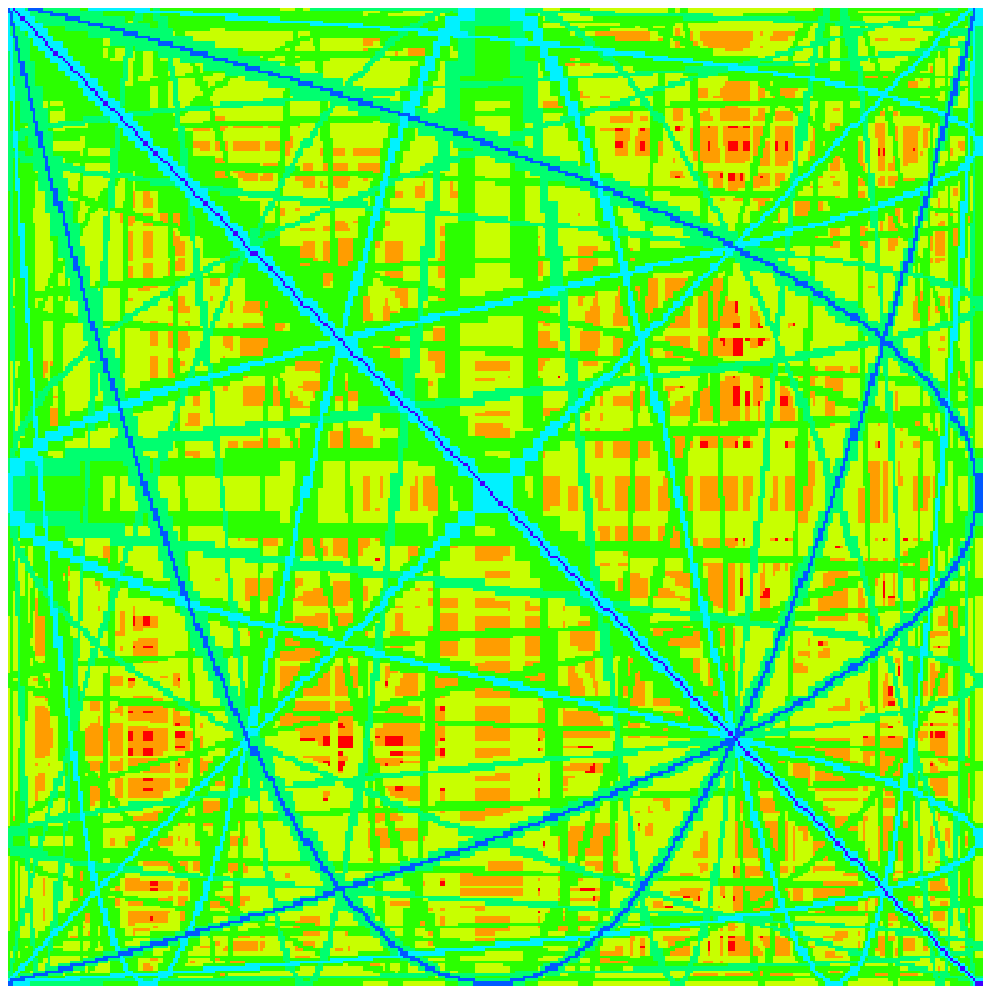,width=0.75\columnwidth,keepaspectratio,clip=}

$ $

\epsfig{figure=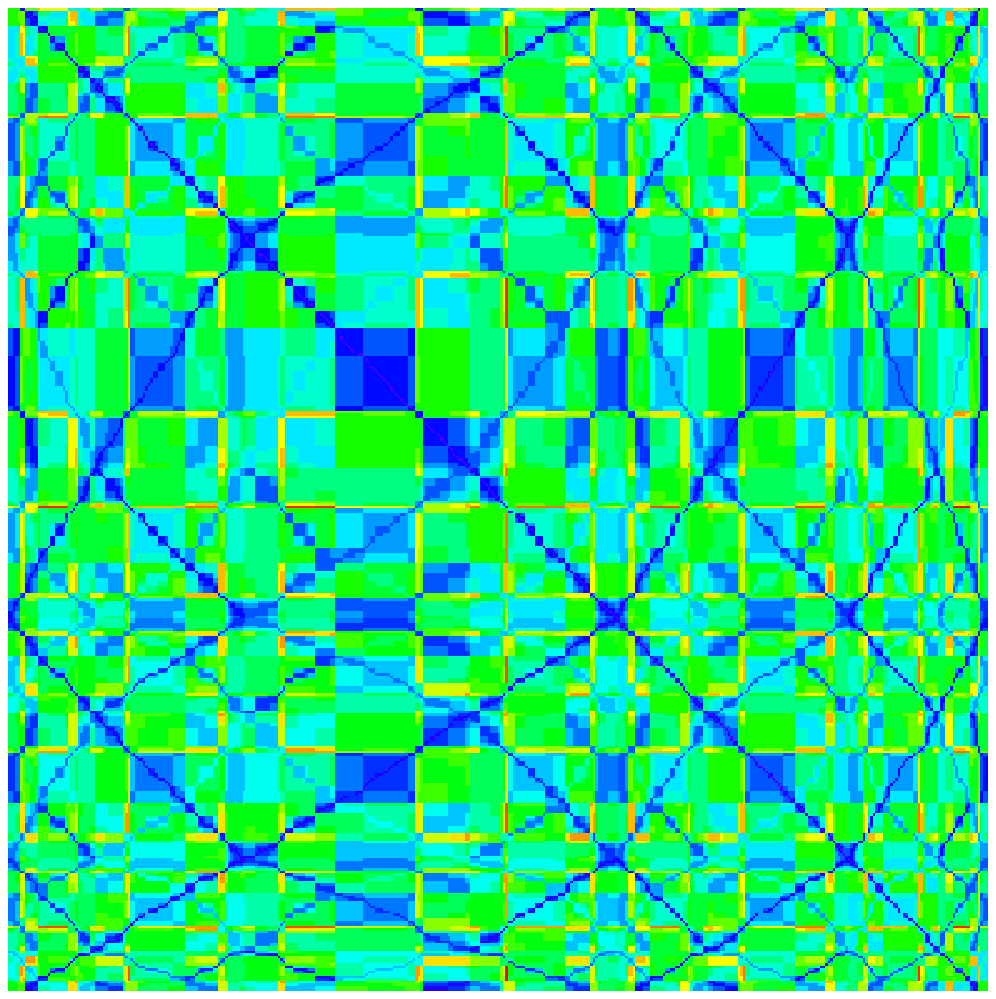,width=0.75\columnwidth,keepaspectratio,clip=}
\end{center}
\caption{Color scale plots (gray scale in paper version) of $\widehat{M}$ 
for $a=2$~(a) and $a=a_c$ (b). Scale ranges from black and blue 
($\ell=0$ and $\ell=1$) to red ($\ell=D$). Number of levels in (a) is much 
smaller than in (b).  Neighborhood structure changes abruptly in the distinct 
regimes.}
\label{fig:colormatrix}
\end{figure}

Finally, we use the information in $\widehat{M}$ to obtain images,
in color (or gray) scale,
of the network neighborhood structure.
They provide a vivid and easy visualization of their distinct
properties of the attractor. For $a=2$, the first order neighborhood is
distributed
along the parabola described by the r.h.s. of (\ref{eq:logistic}) (see 
Fig.~\ref{fig:colormatrix}a). It shows how the second and higher order 
neighborhood evolve according to the higher order iterates of the quadratic map.
However, mixing and finite size effects stemming from a finite
graining smears the higher order iterates. The situation is distinct
for $a_c$, when the attractor is a $d_{F,A}=0.54...$ dust spread out
in the $[-1,1]$ interval. Only boxes containing part of the dust remain
in the network, so that contiguous numbered nodes are not actually neighbors
in phase space. The resulting image (Fig.~\ref{fig:colormatrix}b)
displays a fine and intertwined tessitura that reflects a very peculiar 
behavior of the trajectories at the onset of chaos\cite{rob04,rob05,rob06}.

\section{Conclusion}
\label{sec:conclusions}

In this work we explore the idea of mapping chaotic systems onto complex
networks. Networks are constructed according to a well defined
methodology, and results for the logistic map indicate how their
properties can be associated to those of the attractor in phase space.
Trajectories in distinct dynamical regimes are explored in order to show
how the major differences in phase space are reflected in the networks.
The networks show several features of small-world and scale-free networks,
but they do not fully match with those networks generated by the specific
algorithms described in \cite{watts98,baralb99}. The analysis of the
networks in the neighborhood of $a_c$ reveals that the $N\times W$
dependence, measured by $d_{F,A}$, has a sharp transition at the onset of
chaos. So, the distinct character of the trajectories in phase space is
indeed reflected in the network. However, with exception of $C$, the
results for the other indices ($\langle d \rangle,D$ and $p(k)$) change in
a much smoother way with respect to changes in the parameter $a$.

Acknowledgements: This work was supported by the Brazilian
Agencies CNPq and FAPESB.

\end{document}